\documentclass[12pt]{spieman}  % 12pt font required by SPIE;
\usepackage{amsmath,amsfonts,amssymb}
\usepackage{tocloft}
\usepackage{xcolor}
\usepackage{graphicx}
\usepackage{setspace}
\usepackage{tocloft}
 \usepackage{tabularray}
\usepackage{lineno}
\usepackage{algorithmic}
\usepackage{algorithm,algorithmic}
\usepackage{xfrac}
\usepackage{lipsum}
\usepackage{multirow}
\usepackage [english]{babel}
\usepackage [autostyle, english = american]{csquotes}
\MakeOuterQuote{"}
\usepackage{caption}
\usepackage{ulem}

\title{A sectorial arrangement of spatially-shifted grating for composite Laguerre-Gaussian OAM beam generation}

\author[a,*]{Pratyush Pushkar}
\author[a]{Nirjhar Kumar}

\affil[a]{Indian Institute of Technology Madras, Experimental Optics Laboratory, Department of Electrical Engineering, Chennai, India, 600036}

\cftpagenumbersoff{figure}
\cftpagenumbersoff{table}
\DeclareCaptionLabelFormat{bold}{\textbf{#1 #2}} % Define a new label format
\begin{document} 
\maketitle
\begin{abstract}
The first diffraction order of a planar 1D grating has a plane wavefront for a plane wave incidence. We propose an algorithm to tailor the shape of this wavefront to the desired propagation mode. The algorithm works by segmenting the 1D grating into sectors and shifting gratings in each sector along the grating vector by some fraction of the grating period. The algorithm determines the fractional shift required in each sector to impart a desired shape to the wavefront. Through simulations and experiments, we show the applicability of the design process by demonstrating the coaxial generation of one or multiple Laguerre-Gaussian orbital angular momentum beams. In addition, we demonstrate the independent regulation of relative intensities and phases of components in the generated beam.
\end{abstract}

% Include a list of up to six keywords after the abstract

% Include email contact information for the corresponding author
{\noindent \footnotesize\textbf{*}Pratyush Pushkar,  \linkable{ee16d020@smail.iitm.ac.in} }

\begin{spacing}{1}   % use double spacing for rest of manuscript
\section{Introduction}
\label{sect:intro}  % \label{} allows reference to this section
Laguerre-Gaussian (LG) beams, which carry a helical wavefront and have a doughnut-shaped intensity profile, have been studied over the last three decades. These beams have azimuthally varying phase profiles given by $e^{j\ell\phi}$ and possess an orbital angular momentum (OAM) $L=\ell\hbar$ per photon where $\ell\in \textbf{Z}$ indicates the beams' topological charge and $\phi$ is the azimuthal angle \cite{allen1992orbital}. Thus, these beams are also known as LG-OAM beams. LG beams also have radial modes represented by radial index $p$, where $p\in \textbf{W}$ determines the number of intensity nulls in the radial direction. These beams have played an important role in many applications, such as in micro-manipulations\cite{padgett2011tweezers,otte2020optical,yang2021optical}, imaging \cite{ritsch2017orbital,gozali2017compact}, encryption \cite{ruffato2015spiral}, sensing \cite{pang2021review}, and quantum communication \cite{sit2017high,bouchard2018quantum,hufnagel2020investigation,cozzolino2019orbital}. Several techniques,  such as a pair of cylindrical lenses \cite{beijersbergen1993astigmatic}, spiral phase plates \cite{beijersbergen1994helical}, and q-plates\cite{marrucci2006optical}, have been used conventionally to generate these beams. 

Further, LG beams with distinct pairs of $\ell$ and $p$ values form an orthogonal basis set, the span of which constitutes composite LG beams - a superposition of several LG beams. Composite LG beams have garnered much interest as they can be used as multiplexed and multicasted beams in optical communication, thus enhancing the transmission capacity \cite{wang2012terabit,bozinovic2013terabit,zhu20161,yan2013multicasting}. Additionally, these beams are also used for multiple particle-trapping and their rotation\cite{paterson2001controlled,macdonald2002creation}, determining the spinning frequency of a rotating object  \cite{lavery2013detection}, and electromagnetically induced transparency\cite{hamedi2021ferris}. One of the standard techniques to generate composite LG beams is using an interferometer-based setup \cite{galvez2006composite,franke2007optical}. However, this technique requires a bulky optical setup with precise alignment and is prone to misalignment due to mechanical vibration. An alternative method involves mapping the amplitude-phase profile of composite LG beams into amplitude-phase or, preferably, solely phase-only diffractive optical element (DOE) patterns \cite{szatkowski2020generation}. These mapping algorithms generate patterns that can be implemented on a spatial light modulator (SLM) \cite{zhu2019review} or compact on-chip metamaterials \cite{du2015design}, reducing the number of required optical elements and the need for precise alignment. 
   However, phase-only DOE patterns cannot achieve an ideal phase-only transmittance for a finite number of topological charges, except for a single charge \cite{lin2005collinear}. Iterative algorithms such as Gerchberg–Saxton \cite{gerchberg1982optik} and Adaptive-Additive algorithms \cite{soifer2014iteractive} facilitate finding a phase-only DOE pattern for a closer approximation of the desired ideal transmittance. This approximation can aid in achieving high power efficiency in the desired Laguerre-Gaussian (LG) modes while accepting some errors, such as a small power leakage in some non-desired topological charges.
Several improved versions of these iterative algorithms have been proposed \cite{lin2005collinear,lin2006synthesis,lin2007multiplexing,zhu2015simultaneous}. Due to iteration steps, these methods are inherently slow to update the patterns, and if one or more constituent modes of a composite beam are removed or changed, then all the iteration steps have to be repeated. Moreover, these methods depend strongly on the initial guess for convergence. Various non-iterative mapping algorithms have also been proposed. A common non-iterative approach is to divide the pattern area into different sections, each allocated for generating a specific component of the composite LG beams. A random phase mapping algorithm involves multiple discontinuous sections randomly assigned to a component of the composite LG beams \cite{szatkowski2020generation}. An algorithm for a sliced phase pattern has been proposed, which involves azimuthal windowing of an incident LG beam, resulting in an equivalent spectral leaking of power into multiple LG modes \cite{yan2013multicasting}. Both the proposed algorithms lack the independent intensity control of the constituent modes, whereas the latter is also limited to generating constituent LG modes equispaced in $\ell$. A hybrid fork grating design involves dividing the pattern area into concentric sections, but with the number of concentric sections limited by the grating area and the spectral leakage of power into the multiple unwanted radial LG modes \cite{kumar2021single}. An alternate non-iterative algorithm involves two SLMs to control the phase and amplitude of the composite beam independently. However, this technique is prone to unwanted phase distribution,  necessitating the inclusion of extra optical components between the SLMs to mitigate this issue. Adding extra optical components and the requirement of two SLMs increases the system's complexity \cite{zhu2014arbitrary}.

This work proposes a non-iterative algorithm to design DOE patterns for composite LG beam generation. The DOE pattern is derived from a 1D grating with a circular aperture divided into sectors, and each sector is perturbed by introducing a shift in the grating along the grating vector. The shift in different sectors can be different fractions of the grating period, depending upon the desired phase profile. The obtained sectorally-spatially-shifted grating  (S$_3$G)  has been introduced in Sec. \ref{dm}. For the generation of an LG beam, the convergence of S$_3$G to a fork grating validates our design principle. LG beams in the first diffraction order (1$^{st}$-DO) of the S$_3$Gs were further verified using a Python simulation and experimentally demonstrated using an SLM. In Sec. \ref{dm}, a composite-S$_3$G (c-S$_3$G) is proposed, and in Sec. \ref{rnd}, it is simulated and experimentally demonstrated for the generation of composite LG beams. In addition, we show simultaneous independent control of the intensities and phases of the constituent LG beams. Further, the c-S$_3$G is compliant with incremental updates, i.e., to add, remove or modify one or more constituent LG modes in the 1$^{st}$-DO without altering the entire structure.
    In essence, the proposed algorithm generates a bitmap image that can be used as a mask for an SLM or converted into a graphic data system (GDS) file for fabrication. An SLM was chosen for this study due to its significant advantages in applications requiring dynamic configurability and flexibility. Since no on-chip device was fabricated, the fabrication option was not explored further in this report.

\section{Design Methodology}
\label{dm}
\subsection{Theory and Simulation}
\begin{figure*}[b]
\centering
\includegraphics{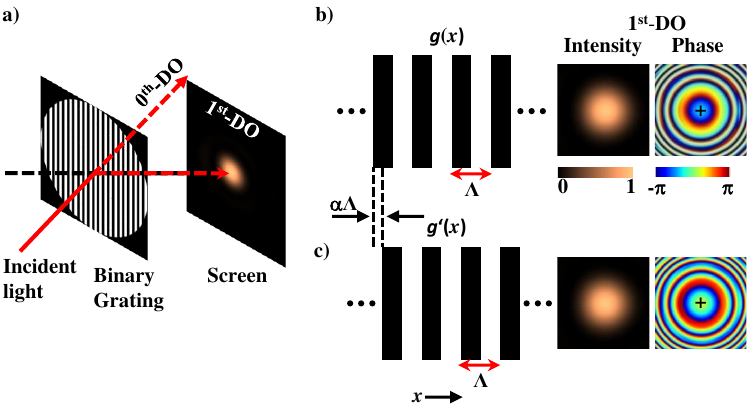}
\caption{(a) An illustration of the optical setup used to simulate the 1$^{st}$-DO of a grating. A zoomed sectional view of a standard 1D grating along with the intensity and phase profiles obtained in the simulations, before (b) and after (c) the grating was spatially shifted by $\alpha\Lambda$ along the grating vector \textit{x}.}
\label{fig01}
\end{figure*}

    Figure \ref{fig01}(a) shows the schematic of the optical setup simulated using Huygens' principle \cite{kumar2021single} with the parameters given in Table \ref{tab1}.
The setup consisted of a 1D binary grating on which a Gaussian beam was incident at an angle such that the transmitted 1$^{st}$-DO was incident normally on a screen. A portion of this grating is shown in Fig. \ref{fig01}(b), along with its simulated 1$^{st}$-DO intensity and phase profile. For the simplicity of the theoretical analysis, this grating can be represented in 1D as a periodic function  $g(x)$ with the value in the principal period \big[-$\frac{\Lambda}{2},\frac{\Lambda}{2}\big]$ given by Eq. (\ref{deqn_ex1}).

\begin{equation}
\label{deqn_ex1}
{g(x)} = \begin{cases}
1,&|x| \leq \Lambda/4 \\ 
0, &|x| > \Lambda/4
\end{cases}
\end{equation}
 The Fourier series (FS) representation of $g(x)$ is provided in Eq. (\ref{gx_in_fourier}), where $C_{\kappa}$ denotes the complex Fourier coefficient, which is non-zero for $\kappa$ = 0 or odd, as dictated by Eq. (\ref{fourierCoff}). The magnitude of $C_{\kappa}$ determines the amplitude, while its phase corresponds to the phase of the $\kappa^{th}$-diffraction order. 
\begin{subequations}\label{gx_in_fourier} 

\begin{equation}
g(x) = \sum_{\kappa=-\infty}^{\infty} C_{\kappa} e^{j\kappa\frac{2\pi}{\Lambda}x}
%\label{gx_in_fourier} 
\tag{\textcolor{black}{2}}
\end{equation}

\begin{equation}
 C_{\kappa} =\frac{1}{\kappa\pi}sin(\frac{\kappa\pi}{2})
  \label{fourierCoff}
\end{equation}
\end{subequations}

 The binary amplitude grating $g(x)$, depicted in Fig. \ref{fig01}(b) when shifted by $\alpha \Lambda$ along the x-axis results in $g^{'}(x)$ shown in Fig. \ref{fig01}(c) which can be expressed as Eq. (\ref{shiftedGrating}). The 1$^{st}$-DO simulated intensity and phase profile of both $g(x)$ and $g^{'}(x)$ are also shown in Fig. \ref{fig01}(b) and Fig. \ref{fig01}(c), respectively. It can be observed that the intensity and the phase profile for $g(x)$ and $g^{'}(x)$  were identical, except for a relative phase shift $\phi$ between them.
The value of $\phi$ can be calculated at any pair of corresponding points in the phase profiles. The phases at the points marked as `+' in  (b) and (c)  were respectively found to be $\approx$ $-$0.34$\pi$ and 0.16$\pi$; thus, $\phi \approx 0.16\pi + 0.34\pi = \pi/2$.   This phase shift can be explained by comparing the FS expansion of $g(x)$  and $g^{'}(x)$. 
 The FS representation of $g^{'}(x)$ using Eq. (\ref{gx_in_fourier}) is given by Eq. (\ref{dashedFS}).

\begin{equation}
\label{shiftedGrating}
g'(x)=g(x-\alpha\Lambda)
\end{equation}
\begin{equation}
\label{dashedFS}
\sum_{\kappa=-\infty}^{\infty} C'_\kappa e^{j\kappa\frac{2\pi}{\Lambda}x}=\sum_{\kappa=-\infty}^{\infty} C_\kappa e^{j\kappa\frac{2\pi}{\Lambda}x} e^{-j\kappa\frac{2\pi}{\Lambda}\alpha\Lambda}
\end{equation}
The Fourier coefficient  $C^{'}_{\kappa}$ of $g^{'}(x)$ is given in terms of $C_{\kappa}$ of $g(x)$ in Eq. (\ref{phase_shifted_coeff}). 
\begin{equation}
\label{phase_shifted_coeff}
    C'_\kappa= C_\kappa e^{-j2\kappa\pi\alpha}
\end{equation}
The phase difference $\phi$ between the phase profiles of $g(x)$ and $g'(x)$ is equal to the phase difference between $C^{'}_{\kappa}$ and $C_{\kappa}$, as given by Eq. (\ref{deqn_ex6}).
\begin{equation}
\label{deqn_ex6}
\phi  = 2 \kappa \pi \alpha 
\end{equation}

For the 1$^{st}$-DO, i.e., $\kappa$ = 1, and $\alpha=0.25$ used in Fig. \ref{fig01}(c), the value of $\phi = \pi/2$  is approximately equal to the value of $\phi$ calculated previously from the phase profile. Hence, a spatial shift of $\alpha \Lambda$ in a grating corresponds to a phase shift of $2\pi\alpha$ in the 1$^{st}$-DO. This was observed to be true for both amplitude and phase gratings.

\begin{figure*}[t]
\centering
\includegraphics{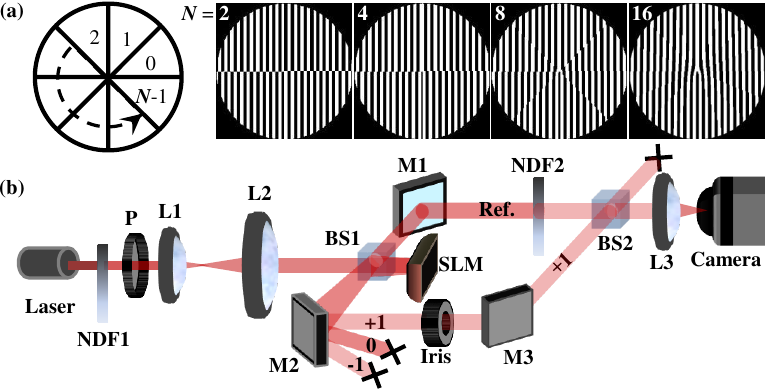}
\caption{(a) A layout of the sectorial division of a grating and illustrations of the gratings, i.e., S$_3$Gs obtained after spatially shifting the gratings in the sectors. The S$_3$Gs illustrated are for the desired phase profile of an $\ell= 1$ and $c=0$ LG beam. It can be observed that as $N$ increases, the S$_3$G evolves into a standard $\ell= 1$ fork grating. (b) An illustration of the experimental setup used to generate the LG and Composite LG beams and validate the tuning of the phase profile in the 1$^{st}$-DO of the S$_3$G. NDF1 and NDF2 are neutral density filters,  P is polarizer, L1 - L3 are lenses with focal lengths of 25 mm, 100mm, and 25 mm, respectively, M1- M3 are mirrors, BS1 and BS2 are beam splitters, and SLM is the spatial light modulator. A 632.8 nm wavelength He-Ne laser was used. The reference beam path Ref. was blocked (unblocked) to capture the camera's 1$^{st}$-DO intensity profile (interferogram) in the camera.}
\label{fig02}
\end{figure*}

The spatial shift can also be introduced in sections of a grating. This can be leveraged to tune the phase profile of the 1$^{st}$-DO by introducing a different spatial shift in different sections of a grating, resulting in a sectorally-spatially-shifted grating (S$_3$G). Figure \ref{fig02}(a) shows sections of a grating as $N$ sectors numbered $n=0$ to $N$-1 with spatial shift $\alpha(n)\Lambda$ in each sector. For a desired azimuthal phase profile $ \phi(\theta)$ in the 1$^{st}$-DO, $\alpha(n)$ of the S$_3$G can be calculated from the discretized  $ \phi(\theta)$, i.e., $\phi(n)$ using Eq. (\ref{deqn_ex6}). $ \phi(\theta)$  and the corresponding $\phi(n)$ for the generation of an LG beam of charge $\ell$ and initial phase $c$ are given by Eq. (\ref{deqn_ex7}) and Eq. (\ref{deqn_ex8}), respectively.

\begin{equation}
\label{deqn_ex7}
\phi(\theta)=\ell\theta + c
\end{equation}
\begin{align}
\label{deqn_ex8}
\phi(n)=\ell n\delta\theta + c \text{, where } \delta\theta = 2\pi/N
\end{align}
Figure \ref{fig02}(a) further shows $\ell= 1$ and $c = 0$ S$_3$Gs for different values of $N$. For $\ell=1$, an edge-dislocation manifests for $N \geq 4$, whereas for $N\geq16$, it evolves into a well-known structure for LG beam generation: a fork grating. These gratings were simulated using the simulation setup shown in Fig. \ref{fig01}(a) and were experimentally verified using the experimental setup shown in Fig. \ref{fig02}(b). The parameters used for the simulation and experiment are shown in Table \ref{tab1}. Section \ref{rnd} discusses the difference in the parameter taken. The generic algorithm for the S$_3$G design is given in Algorithm \ref{alg:S$_3$G}.

\begin{algorithm}
\caption{Algorithm for a S$_3$G design}\label{alg:S$_3$G}
 \begin{algorithmic}[1]
 \renewcommand{\algorithmicrequire}{\textbf{Input:}}
 \REQUIRE Desired transverse phase profile, number of sectors (\textit{N}), and grating period ($\Lambda$)
 %\ENSURE  out
 \\ \textit{Initialisation} : Design a 1D grating with  period $\Lambda$.
  \STATE Quantize the desired phase profile into \textit{N} levels. \\
  \hspace*{0.5cm}% 
 \begin{minipage}{0.83\textwidth}%
 \vspace{1 mm}
 \textit{Corollary: $\Rightarrow$ The desired phase profile is divided into M constant phase sections, numbered m = 0 to M-1, each with a constant phase $\phi(m)$. (For the case of a spiral phase profile: M = N, sections = sectors).}
\end{minipage}%
\vspace{1 mm}
\\  \STATE Divide the 1D grating into identical \textit{M} sections.\\
  \hspace*{0.5cm}%
   \begin{minipage}{.83\textwidth}%
  \vspace{1 mm}
\textit{For the case of spiral phase profile: Divide the 1D grating into \textit{N} sectors as shown in Fig. \ref{fig02} (a).}
 \end{minipage}%
  \STATE Shift the grating in each section by $\frac{\phi(m)}{2\pi}$ times $\Lambda$ along the grating vector.
 \renewcommand{\algorithmicensure}{\textbf{Output:}}
 \ENSURE  A S$_3$G whose 1$^{st}$-DO has the desired phase profile. 
 \end{algorithmic} 
 \end{algorithm}
 \subsection{Experimental Setup}
\
Figure \ref{fig02}(b) shows the experimental setup used to demonstrate the generation of desired phase profiles in the 1$^{st}$-DO of S$_3$Gs. The setup comprised a 6 mm $\times$ 6 mm sectorial grating of  $\Lambda = $ 96 $\mu$m loaded on an 8-bit PLUTO-2-VIS-016 phase-only SLM. A 6 mm diameter, a horizontally polarised collimated laser beam of wavelength $\lambda =632.8$ nm was incident normally on the SLM. The 1$^{st}$-DO reflected was isolated using a beam splitter and an iris diaphragm and captured on the camera. The phase profile of the 1$^{st}$-DO was inferred from the interferogram obtained by unblocking the reference beam path. 
\begin{table}
\centering\caption{Parameters used in simulation and experiment}
\label{tab1}
\begin{tabular}{|l|r|r|}
\hline
\multicolumn{1}{|c|}{ Parameters used:}        & \multicolumn{1}{c|}{in simulation} & \multicolumn{1}{c|}{in experiment} \\ \hline
Wavelength $\lambda$                            & 632.8 nm                        & 632.8  nm                       \\ \hline
\multirow{2}{*}{Grating size}           & 20.48 $\mu$m $\times$ 20.48 $\mu$m               & 6 mm $\times$ 6 mm                 \\ \cline{2-3} 
                                        & 512 $\times$ 512 pixels                & 750 $\times$ 750 pixels               \\ \hline
Period $\Lambda$                                & 0.96 $\mu$m                          & 96 $\mu$m                           \\ \hline
Beam waist $\omega_0$                           & 10.24 $\mu$m                               & 3 mm                               \\ \hline
Rayleigh Range $z_R$                       & 0.52 mm                        & 44.68 m                      \\ \hline
\multirow{2}{*}{Propagation distance D} & 0.50 mm                        & $\approx$1 m                       \\ \cline{2-3} 
                                        & $\approx z_R$                          & $\approx\sfrac{z_R}{50}$                    \\ \hline
\end{tabular}
\end{table}

\begin{figure*}[t]
\centering
\includegraphics[scale = 0.95]{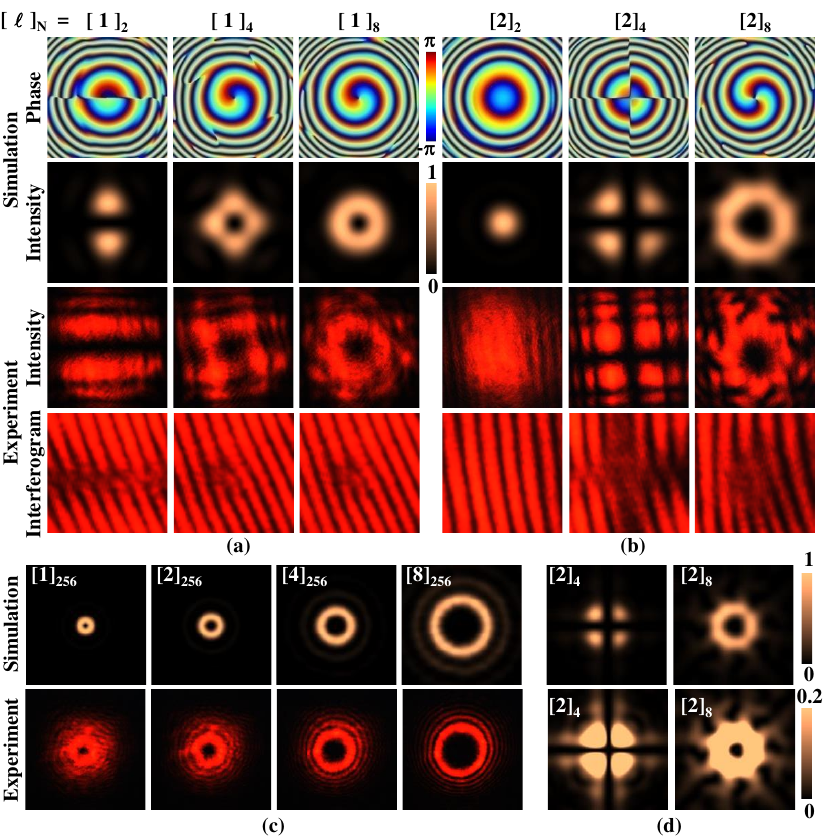}
\caption{ The intensity and the phase/interferogram profiles of the simulated and experimentally obtained $1^{st}$-DOs. (a) and (b) shows the evolution of the profiles with $N$  for $\ell=1$,  $c=0$, and $\ell=2$, $c=0$ respectively. (c) shows the generation of different LG beams using S$_3$Gs, which are fork gratings of respective charges for $N=256$. (d) comparison of simulated intensity profiles for $\ell=2$ with $N=4$ and $N=8$ at different exposure settings.
}
\label{fig03}
\end{figure*}

  \subsection{Choice of N}
Figure \ref{fig02}(a) illustrates fork gratings as a special case of S$_3$Gs. As $N$ increases, an S$_3$G evolves from a 1D grating to a fork grating, causing the corresponding 1$^{st}$-DO evolving from a Gaussian beam to an LG beam. This evolution has been studied through 
simulation and experiments in Fig. \ref{fig03}(a) for $\ell=1$ and Fig. \ref{fig03}(b) for $\ell=2$. With increasing values of $N$, the simulated intensity profile gradually transforms into a doughnut shape, revealing a spiral phase at the center of the 
phase profile for $N \geq 4\ell$. Similarly, experimental results show a comparable 
evolution of the intensity profile with $N$, displaying a fork pattern at the center of the 
interferogram for $N \geq 4\ell$. Thus, lower values of $N$ result in a higher distortion from an ideal LG beam profile, while higher values of $N$ lead to an improvement in beam purity. Additionally, some disparities between the simulated and experimentally obtained intensity profiles can be observed. Additional side lobes/fringes in the experimental results are attributed to higher-order modes stemming from beam impurity. The difference in the visibility of these features in simulated and experimentally obtained intensity profiles can be attributed to differences in higher-order mode visibility in simulation and experiments. In the simulation, the increase in LG beam size with $\ell$ was proportional to ${\ell+1}$, whereas in the experiment, it was proportional to $\sqrt{\ell+1}$\cite{kumar2021single,padgett2015divergence}. Thus, the relative size of the higher-order modes compared to a lower-order mode was larger in the simulation than in the experiment, as inferred from Fig. \ref{fig03}(c). In the simulation, despite similar spectral power leakage, the larger relative size of the higher-order mode results in its diminished relative intensity and subsequent lower visibility compared to experimentation. This hypothesis is supported by Fig. \ref{fig03}(d), which shows the overexposed simulated intensity profile, revealing the presence of these extra side lobes/fringes. The difference in orientation of the simulated and experimentally obtained intensity profiles is addressed later in the paper.
 
 To avoid distortion, a sufficiently high value of $N = 256$ was used throughout this work unless stated otherwise. However, even with this value of $N$ in Fig. \ref{fig03}(c), sidelobes are evident, potentially attributed to the windowing effect caused by the finite grating aperture \cite{kumar2021single}. Once more, differences in the visibility of these sidelobes between experiment and simulation, as discussed earlier, are apparent.
\subsection{Proposed Method}
\begin{figure}[t]
\centering
\includegraphics[scale = 1]{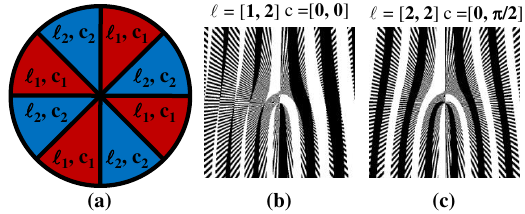}
\caption{(a) A layout of a c-S$_3$G where the spatial shift in sectors alternately corresponds to different desired phase profiles. The odd sectors in red correspond to a phase profile $\phi_1$ of an LG beam of charge $\ell_1$ and initial-phase $c_1$, whereas the even sectors in blue correspond to $\phi_2$ of an LG beam of charge $\ell_2$ and initial-phase $c_2$. (b) and (c) shows the zoomed sectional views of the c-S$_3$Gs obtained for  $\ell=[1$, 2], $c=[0$, 0] and  $\ell=[2$, 2], $c=[0$, $\pi/2$], respectively. }
\label{fig04}
\end{figure}
Instead of assigning all $N$ sectors of an S$_3$G to a single $\phi(n)$, the sectors can be distributed among multiple desired phase profiles $\phi_1(n)$, $\phi_2(n)$, ..., $\phi_K(n)$ to generate a coaxial superposition of phase structured beams. Thus, these composite-S$_3$Gs (c-S$_3$Gs) can generate composite LG beams, a coaxial superposition of two or more LG beams. Fig. \ref{fig04}(a) shows the round-robin distribution of sectors between two LG beam phase profiles, $\phi_1(n)$ and $\phi_2(n)$ given by Eq. (\ref{deqn_ex14}) and Eq. (\ref{deqn_ex15}), respectively, in a c-S$_3$G.

\begin{equation}
\label{deqn_ex14}
\phi_1(n)=\ell_1n\delta\theta + c_1
\end{equation}

\begin{equation}
\label{deqn_ex15}
\phi_2(n)=\ell_2n\delta\theta + c_2
\end{equation}
The resultant $\phi(n)$ is given by Eq. (\ref{deqn_ex16})
\begin{equation}
\label{deqn_ex16}
\phi(n) = \begin{cases}
\phi_1(n),& \text{if n is odd}\\
\phi_2(n), & \text{if n is even}
\end{cases}
\end{equation}
Figure \ref{fig04}(b) and (c) show a section of c-S$_3$Gs obtained for two distinct sets of $\ell$ and $c$ values. The generic algorithm for the c-S$_3$G design is given in Algorithm \ref{alg:cS$_3$G}.
\begin{algorithm}
 \label{cS$_3$G}
 \caption{Algorithm for c-S$_3$G design}\label{alg:cS$_3$G}
 \begin{algorithmic}[1]
 \renewcommand{\algorithmicrequire}{\textbf{Input:}}
 \REQUIRE $\ell = [\ell_1, \ell_2, ... , \ell_K]$, $c = [c_1, c_2, ... , c_K]$, N, and $\Lambda$
 %\ENSURE  out
 \\ \textit{Initialisation} : Design a 1D grating with a circular aperture and period $\Lambda$.
 \STATE Divide the 1D grating into N sectors.
 \\ \textit{LOOP Process}
  \FOR {$n = 0$ to N-1}
  \STATE $k = (n \mod K) +1$
  \STATE $\phi = \ell_k \times 2\pi n/N + c_k$
  \STATE Shift the grating in the sector $n$ by $\frac{\phi}{2\pi} \times \Lambda$ along the grating vector.
  \ENDFOR
 \renewcommand{\algorithmicensure}{\textbf{Output:}}
 \ENSURE  A c-S$_3$G whose 1$^{st}$-DO has the desired composite-LG beam. 
 \end{algorithmic} 
 \end{algorithm}
\section{Result and Discussion}
\label{rnd}
\begin{figure*}[!t]
\centering
\includegraphics[scale = 1]{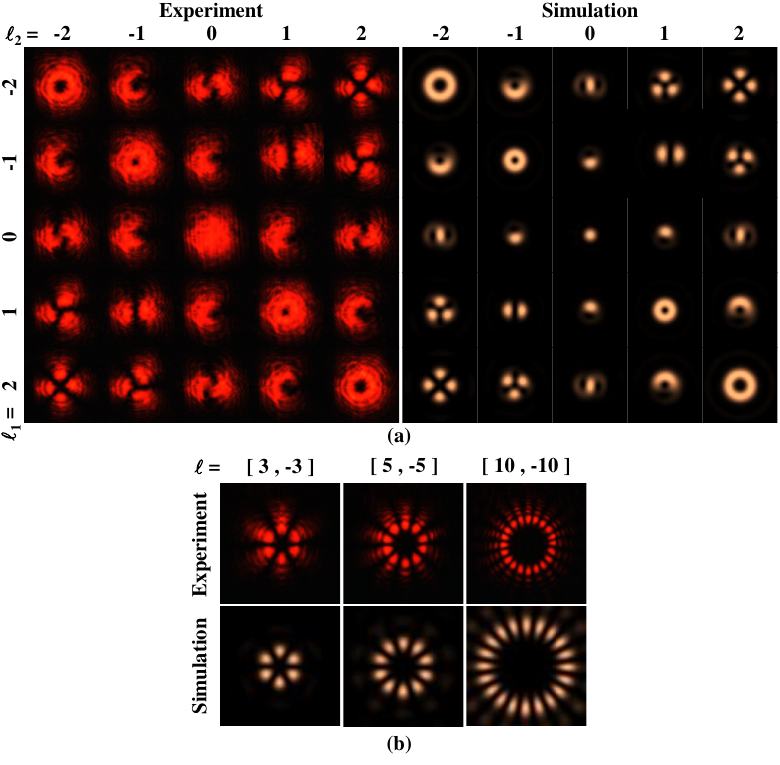}
\caption{Intensity profiles of composite beams obtained using c-S$_3$Gs with in-phase combinations ($c = [0, 0]$) of different $\ell$. (a) $\ell=$ [ $\ell_1$, $\ell_2$] where $\ell_1 , \ell_2 \in  \textbf{Z} \cap$ \{-2 to 2\} (b) $\ell=$ [3, -3], [5, -5], and [10, -10]. }
\label{fig05}
\end{figure*}
Figure \ref{fig05}(a) shows a collage of composite beams generated using the c-S$_3$Gs with zero initial phases and different combinations of $\ell_1$ and $\ell_2$. The experimental and simulated intensity profiles can be observed to have one or more bright regions separated by intensity nulls, resulting in the formation of petal-shaped patterns where the number of petals is given by $|\ell_1-\ell_2|$ \cite{kumar2021single}. For instance, when $\ell_1 = 2$ and $\ell_2 = -2$, the number of petals is 4. However, along the principal diagonal of the collage, where $\ell_1 = \ell_2$, no petal patterns can be observed, and the shape of the beams is that of LG beams. With the beam propagation, the petal patterns were observed to diverge and rotate about the center \cite{huang2016composite}. 
   The beam divergence depends on $z_R$, whereas the rotation depends upon the Guoy phase $\psi_{\ell}(z)$ of the LG beam. $\psi_{\ell}(z)$ is given as\cite{huang2016composite}:
\begin{equation}
\label{deqn_ex17}
\psi_{\ell}(z)=(\ell+1)tan^{-1}(\frac{z}{z_R})
\end{equation}
For composite LG beams with charges $\ell_1$ and $\ell_2$, the relative Guoy phase $\Delta \psi_{\ell_1,\ell_2}(z)$ between the component modes can be expressed as:
\begin{equation}
\label{deqn_ex18}
\Delta \psi_{\ell_1,\ell_2}(z)=(\ell_1-\ell_2)tan^{-1}(\frac{z}{z_R})
\end{equation}

The orientation of the captured composite LG beam depends on $\Delta \psi_{\ell_1,\ell_2}(z=D)$. The discrepancy in orientation arises from the difference in the value of $D=z_R$  used in the simulation compared to  $D=z_R/50$ used in the experiment.

Further, the difference in the relative sizes of the LG beams in the experiment and simulation was similar, as observed earlier in Fig. \ref{fig03}(c). It was observed that the size of the LG beam increases with $\ell$. In experiments, this increase was proportional to $\sqrt{\ell+1}$, while in simulations, it followed a proportionality to $\ell+1$ \cite{kumar2021single,padgett2015divergence}.
This is evident from the principal diagonal in Fig. \ref{fig05}(a). The difference in the relative sizes resulted in the difference in the relative intensities. The peak intensity in the experiment was almost independent of charge; however, in simulation, it was observed to decrease with an increase in $\ell$. This variation in intensity can not be observed in the simulated collage or Fig. \ref{fig03}(c), as the intensity profiles were self-normalized for better visibility. Figure \ref{fig05}(b) shows the generation of flower modes of different charges $\ell$, i.e., composite LG beams with charge -$\ell$ and + $\ell$ \cite{li2015measuring}. 
    The disparities in $\Lambda$ and D values between the simulation and experiments, as evident in Table \ref{tab1}, stem from the need to address limitations in the Huygens' simulator. This simulator considers each grating pixel as a point source emitting a spherical wave \cite{kumar2021single}, introducing an additional spherical phase in the diffraction order when a plane or Gaussian beam illuminates the grating. With increasing D, this spherical phase gradually overwhelms the c-S$_3$G phase, leading to compromised accuracy. D was intentionally set lower in the simulation than in the experiment to mitigate this effect in simulation. However, this adjustment presents challenges in isolating the desired diffraction order, necessitating a reduction in $\Lambda$. Hence, the simulation adopted a scaled-down $\Lambda$, reduced to $1/100^{th}$ of that used in the experiment, as noted in Table \ref{tab1}. This scaling ensures comparability with the incident beam wavelength, thus ensuring the 1$^{st}$-DO is well separated from the 0$^{th}$-DO.

\begin{figure}[b]
\centering
\includegraphics[scale = 1]{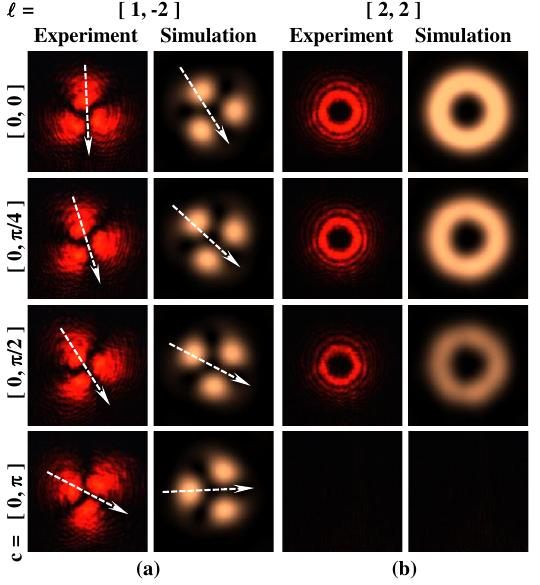}
\caption{Effect of relative initial phase $\Delta c$, i.e., $c = [0, \Delta c]$ on the intensity profile for (a) $\ell_1 \neq \ell_2$ (b) $\ell_1 = \ell_2$. Dashed arrows show the rotation of the intensity profile about the central null with the variation in $\Delta c$. }
\label{fig06}
\end{figure}

Figure \ref{fig06} shows the effect of the initial phases $c_1$ and $c_2$. It was observed through simulation and experiment that the intensity profile of the generated composite beam undergoes rotation with variations in $c_1$ and $c_2$. This rotation depended on the relative phase value $\Delta c = c_1 - c_2$ rather than on the individual values of $c_1$ and $c_2$. For $\ell_1 \neq \ell_2$, a 0 to 2$\pi$ variation in $\Delta c$ resulted in the rotation of the composite beam by 2$\pi/(\ell_1-\ell_2)$ radians, and the sign of ($\ell_1-\ell_2$) determining the sense of rotation. Figure \ref{fig06}(a) shows the rotation of the intensity profile with $\Delta c$ for $\ell_1 = 1$ and $\ell_2 = -2$. However, for $\ell_1 = \ell_2$, a 0 to 2$\pi$ variation in $\Delta c$ resulted in the variation of the peak intensity of the generated LG beam from a maximum at $\Delta c$ = 0 to zero at $\Delta c$ = $\pi$ and back to the maximum at $\Delta c$ = 2$\pi$. This was due to the interference between LG beams generated from the even and the odd sectors of the grating, which was fully destructive for $\Delta c = \pi$ and fully constructive for $\Delta c = 0$ and 2$\pi$. Figure \ref {fig06}(b) shows the effect of $\Delta c$ for $\ell_1 = \ell_2 = 2$.

\begin{figure}[t]
\centering
\includegraphics[scale = 1]{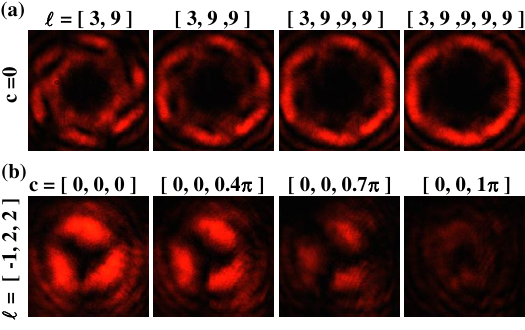}
\caption{ Adjusting the proportion of the charges in a composite LG beam. (a) Coarse adjustment: increasing the proportion of charge $\ell_2$ by increasing its round-robin weight. (b) Fine adjustment: increasing the proportion of charge $\ell_1$ by destructive interference between the sectors with charge $\ell_2$.}
\label{fig07}
\end{figure}
Further, the effect of adjusting the proportion of charges by a weighted round-robin distribution method was studied. It was observed that with the increase in the proportion of charge $\ell_2$, the intensity profile of a [$\ell_1$, $\ell_2$] composite beam evolves into the intensity profile of an $\ell_2$ charge LG beam. Figure \ref{fig07}(a) shows the effect of increasing the proportion of charge 9 in a [3, 9] composite LG beam by varying the round-robin weight ratio. It can be observed that the visibility of the petals/null gradually reduces from a six-petal composite beam for the 1:1 weight ratio to a no-petal LG beam for the 1:4 weight ratio. This method can be extended to obtain intensity control over multiple constituent modes of a composite LG  beam. Figure \ref{fig07}(b) shows the effect of increasing the proportion of charge -1 in the [-1, 2] composite LG beam by the destructive interference of the other charge with itself (as discussed in Fig. \ref{fig06} (b)). Both methods (Fig. \ref{fig07}(a,b)) can be used to control the relative intensities of the constituent modes in the composite LG beam. However, due to destructive interference in (b), method (a) offers comparatively higher diffraction efficiency. On the other hand, method (b) provides a relatively fine intensity control.  As discussed before, a single phase element has an inherent limitation: when the phase profile is structured, the amplitude profile also changes. S$_3$G leverages this limitation to control the amplitude profile of the generated beam in a desired manner (to some extent) by adjusting the phase profile using a 1D grating.
\begin{figure}[t]
\centering
\includegraphics[scale = 1]{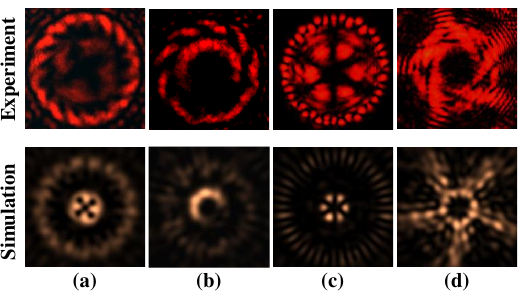}
\caption{Intensity profiles of the composite beams obtained using c-S$_3$Gs with sectors divided among three or more charges. (a) $\ell=$ [-3, 14, 14, 1, 14,
14, -3]$_{280}$, $c=$[0.8$\pi$, $\pi$, $\pi$, 1.6$\pi$, $\pi$, $\pi$, 0.8$\pi$]$_{280}$ (b) $\ell=$[13, 13, 4, 13, 4, 3, 3]$_{280}$,  $c=$[0.8$\pi$, 0.8$\pi$, 0.6$\pi$, 0.8$\pi$, 0.6$\pi$, 0.3$\pi$, 0.3$\pi$]$_{280}$ (c) $\ell=$ [3, -3, 20, -20, 20, -20]$_{360}$, $c=$[0, 0, 0, 0, 0, 0]$_{360}$ (d) $\ell=$[5, 10, 15, 20]$_{280}$, $c=$[0.4$\pi$, 0.5$\pi$, 0.2$\pi$, 0.1$\pi$]$_{280}$.}
\label{fig08}
\end{figure}
Finally, composite beams of three or more charges in different proportions and initial phases were generated, as shown in Fig. \ref{fig08}. The value of $N$ was adjusted as a multiple of twice the sum of the round-robin weights to maintain the point symmetry of the c-S$_3$G about the origin. The generated beams in the experiment matched reasonably with the simulation, except for mismatches due to the relative sizes, intensity, and     orientation of the composite LG beams, 
as discussed in Fig. \ref{fig05}.

    Thus, this method can generate composite LG beams with optimized intensity and phase. Furthermore, in scenarios where an SLM cannot be used, the bitmap image generated from the associated algorithm can be converted to a GDS layout to fabricate on-chip devices for composite LG beam generation. However, this comes at the expense of dynamic reconfigurability. In our future work, we plan to study the properties, uses, and limitations of on-chip S$_3$G devices.

\section{Conclusion}
We demonstrated that a spatial shift in the grating corresponds to a phase shift in the DO. By dividing a grating into multiple sectors, each with a different spatial shift, we create phase-structured light in the 1st-DO. We proposed an algorithm that generates a bitmap image, which can be used as a mask for an SLM or converted into a GDS file for fabricating a compact, integrable on-chip device for composite LG beam generation.We validated the design through simulations and experiments using an SLM by successfully generating various LG and composite LG beams. Additionally, we demonstrated the flexibility of our design method in controlling the relative intensity and phase of one or multiple components of a composite LG beam.
% \disclosures 
\subsection*{Disclosures}
The authors declare no conflicts of interest.

\subsection* {Code, Data, and Materials Availability} 
Data in support of the results presented in this paper are available within the article. 
\subsection* {Acknowledgments}
The authors express their gratitude to Prof. Ananth Krishnan for his fruitful discussions on the manuscript. The authors also thank Prof. Shanti Bhattacharya for providing the SLM and acknowledge Mr. Bagath Chandraprasad T and Mr. Jerin Geogy George for their assistance in utilizing the SLM.
%%%%% References %%%%%

\bibliography{report}   % bibliography data in report.bib
\bibliographystyle{spiejour}   % makes bibtex use spiejour.bst

\listoffigures

\listoftables

\end{spacing}
\end{document}